# RECENT RESULTS ON THE TOP QUARK FROM THE TEVATRON COLLIDER EXPERIMENTS

**A.P. Heinson**
Department of Physics, University of California,
Riverside, CA 92521–0413, USA

for the DØ and CDF Collaborations

## Abstract

This paper presents the latest measurements of the properties of the top quark as determined by the DØ and CDF collaborations at the Fermilab Tevatron $p\bar{p}$ Collider. Both experiments have studied the all-hadronic decay mode of $t\bar{t}$ events. The top quark mass measurement has been refined over the past year, with the addition of new channels and a re-evaluation of some of the earlier ones. There are recent studies of the mass of the $t\bar{t}$ system, the helicity of the $W$ bosons in the top quark decays, and of the correlation of the spins of the top and antitop quarks. New results are also reported on searches for charged Higgs bosons in top quark decay, and for electroweak production of single top quarks.

 ann.heinson@ucr.edu

## 1. Status of the Top Quark Analyses

Over the past two years, there has been considerable activity in the publication of measured top quark properties, as several long-term analyses have concluded. Figure 1 illustrates this activity. The start of each arrow on the plot represents the submission of a paper to Physical Review Letters (PRL) or Physical Review D (PRD), and the end of each arrow denotes publication of that paper. The narrow arrows refer to the PRL summary papers and the broad ones represent documentation of the details of the analyses in PRD. Thirteen papers have been published, and five more submitted, with perhaps another dozen to come over the next year.

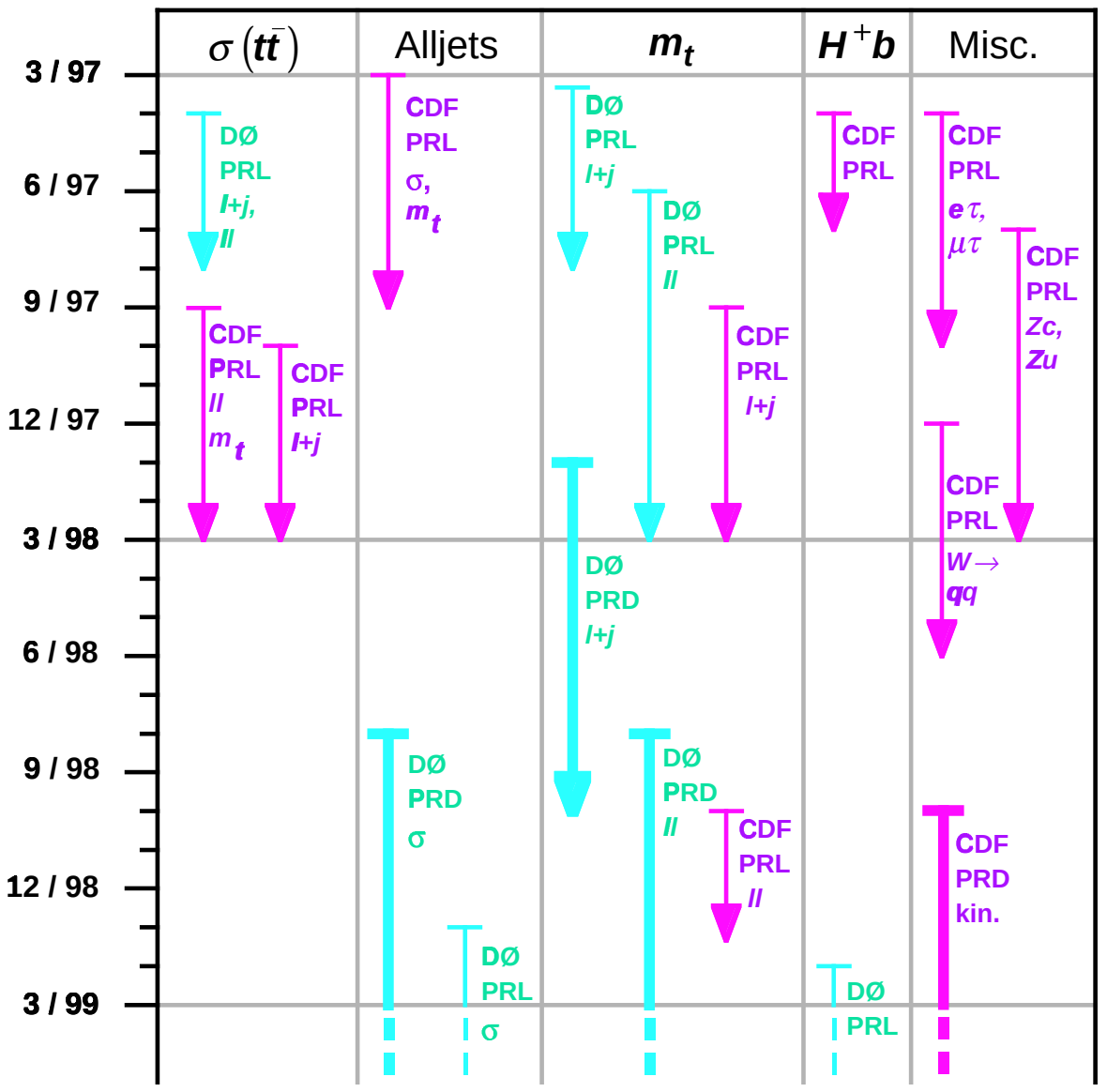

Figure 1 Recent publications on top quark physics from the DØ and CDF collaborations.

DØ has concluded a search for $t\bar{t}$ production in the all-hadronic decay mode. This neural network analysis has led to a new measurement of the $t\bar{t}$ cross section. CDF has made a new measurement of the top quark mass using eight dilepton events, and they have re-evaluated their previous mass measurement using all-hadronic events, incorporating improvements in understanding from the more recent lepton+jets measurement. These results have been combined with DØ's to produce an official Tevatron top quark mass. CDF has measuring the helicity of the $W$ boson from the decay of the top quark, which is sensitive to the top quark decay mechanism. DØ has studied the helicity of the $t\bar{t}$ system itself, by looking for a correlation between the spins of the top and antitop in dilepton events. CDF has optimized event selection for determining the $t\bar{t}$ invariant mass spectrum in a search for high mass resonances. These analyses have not yet been submitted for publication. Based on new theoretical input, DØ has completed a search for the decay of the top quark into a charged Higgs boson, and CDF has updated an old result in this channel, based on new theoretical input. Finally, CDF has two new results from searches for electroweak production of single top quarks which have also not yet been submitted for publication.

## 2. All-Hadronic Decay of $t\bar{t}$ Pairs

First, we review the published CDF measurement[1)] of the all-hadronic decay mode of top quark pairs. They choose events with at least five jets, and total summed scalar transverse energy $H_T$ of the jets > 300 GeV. They then require two jets identified as coming from the decay of a $b$ hadron, or they demand one tagged $b$-jet, together with two kinematic requirements: $H_T$ greater than 75% of the invariant mass of the jets in the final state $\sqrt{\hat{s}}$; and the sum of aplanarity and 0.25% $H_T^{\text{jet}3+}$ greater than 0.54. They measure the background in these samples using QCD multijet events, multiplied by a set of tag rate functions, and obtain the following result:

$$\text{CDF} \qquad \sigma\left(p\bar{p} \to t\bar{t}\right) = 10.1 \pm 1.9^{+4.1}_{-3.1} \text{ pb} \quad (\text{for } m_t = 175 \text{ GeV})$$

In this and other results quoted in the paper, the first error is statistical and the second systematic.

Applying simple selection cuts, and relying on the high $b$-tagging efficiency (~46%) of their precision silicon microstrip vertex detector (SVX), CDF obtains a significance of signal over background of better than three standard deviations.

Since DØ did not have a silicon vertex detector in Run 1, and to obtain a comparably significant observation, they developed a more sensitive event selection technique than simply applying cuts[2]. They devised 10 kinematic variables which show a significant difference between signal ($p\bar{p} \to t\bar{t} \to \geq 6$ jets) and background ($p\bar{p} \to \geq 6$ jets) before including any $b$-tagging information. These 10 variables were used in a neural network trained on an independent sample of multijet data, and on HERWIG[3] $t\bar{t}$ Monte Carlo events, and the output was combined in a second network together with three variables designed to best describe orthogonal qualities expected in $t\bar{t}$ events. The variables are shown in Fig. 2.

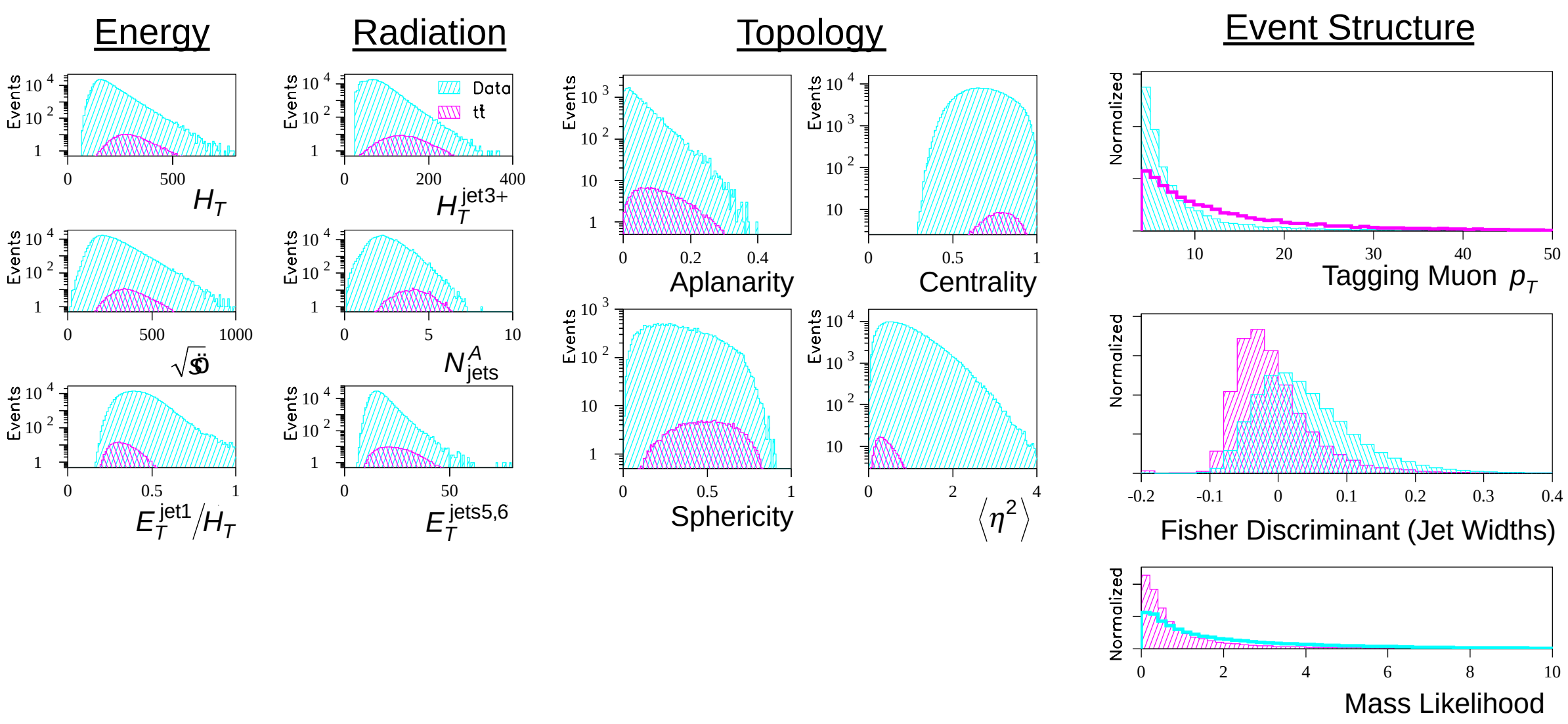

Figure 2 Neural network variables for the DØ $t\bar{t} \to$ alljets analysis. The first 10 variables are used in one network, and the output from that network is used together with the last three variables in a second network.

Most of these variables are self-explanatory; $N_{\rm jets}^{A}$ is the number of jets averaged over a range of $E_T$ thresholds, and weighted by the threshold, so as to take into account the hardness of the jets. The second set of variables consists of the $p_T$ of the tagging muon (harder in the decays of $b$ quarks from top than from $b$ quarks from gluons), a Fisher discriminant constructed to describe the widths of the jets (quark jets are narrower than gluon jets), and a mass likelihood (events that have two dijet combinations near the mass of the $W$, and two sets of 3-jet mass combinations close to each other in value have a likelihood nearer zero).

Figure 3 shows the output of the second neural network, with top-like events (expected signal) clustering near 1.0 and background-like events near 0.0. When a cut is made on the output of this network at 0.94, 18 events remain in the data, with an expected background of $6.9 \pm 0.9$ events, giving a significance for the excess of $3.2\sigma$. Interpolation of the results from nearby top quark mass values gives the following cross section measurement:

DØ $\quad \sigma(p\bar{p} \to t\bar{t}) = 7.1 \pm 2.8 \pm 1.5$ pb $\quad$ (for $m_t = 172.1$ GeV)

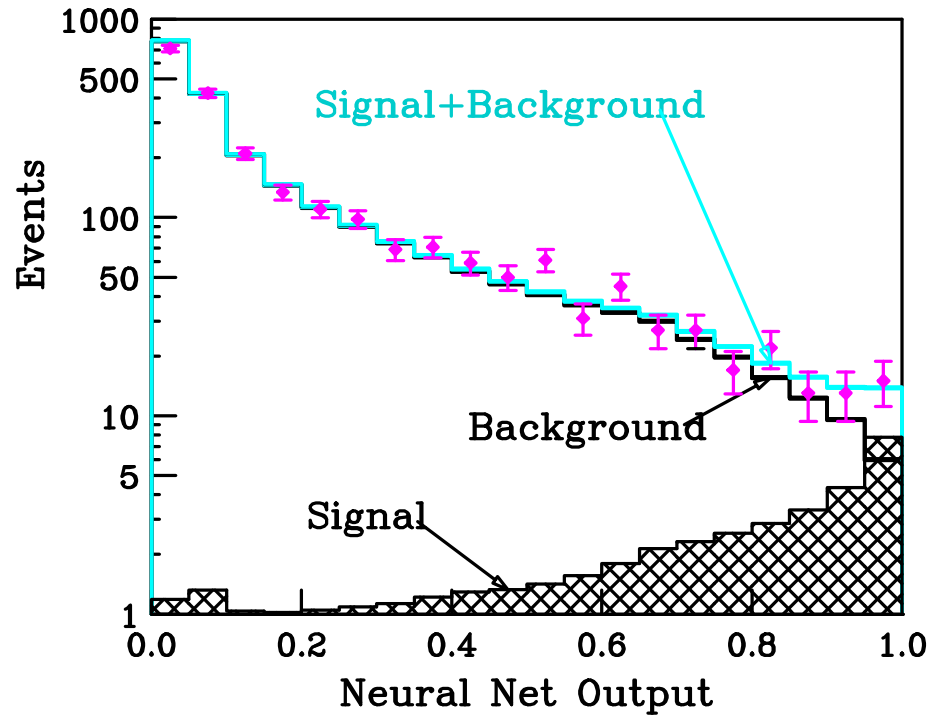

Figure 3 The output distribution for $t\bar{t} \rightarrow$ alljets signal, background, and data, from the second neural network in the DØ analysis.

## 3. Top Quark Mass

Within the past year, CDF has made a new measurement of the top quark mass in the dilepton decay channel[4] by applying the Kondo-Dalitz-Goldstein method,[5] with the choice of variable inspired by DØ's neutrino pseudorapidity weighting method,[6] and re-evaluated the systematic error on their measurement in the all-hadronic decay channel taking into account the method used in the lepton+jets channel.[4] The two collaborations have formed a working group that combined all the latest values to obtain a global Tevatron result.[7] The top quark now has the best measured mass of any quark, with an uncertainty of only 2.9%.

The dilepton mass measurements are illustrated in Fig. 4. DØ's results were published a year ago, and use six events. CDF's recently published result uses eight events, and therefore has a smaller statistical error. The results are:

| | |
|---|---|
| DØ | $m_t = 168.4 \pm 12.3 \pm 3.6$ GeV |
| CDF | $m_t = 167.4 \pm 10.3 \pm 4.8$ GeV |

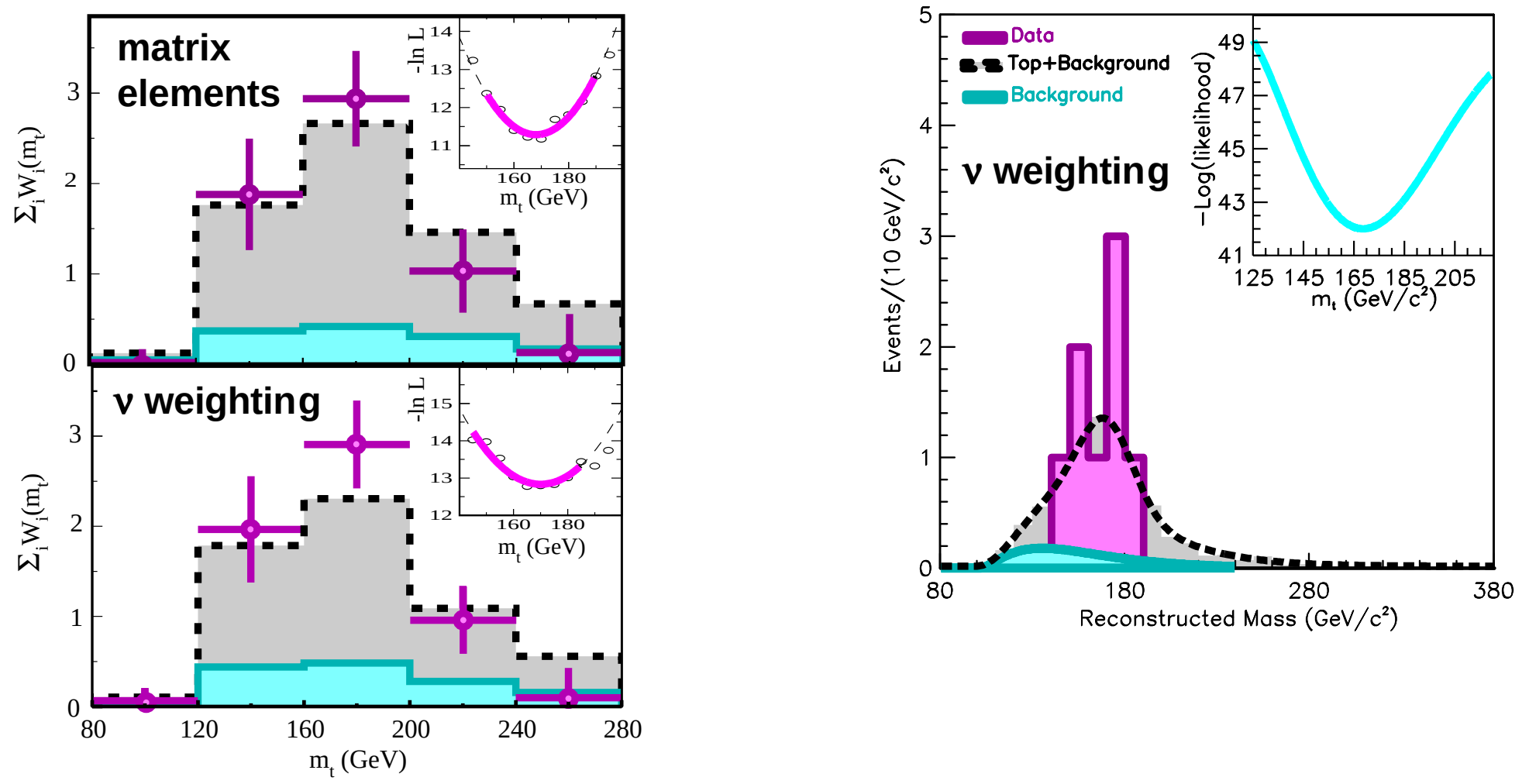

Figure 4 Top quark mass measurement using dilepton events from DØ (left plots) and CDF (right plot).

Figure 5 illustrates the five top quark mass measurements included in the combined mass. It also shows the previous CDF dilepton result obtained from the distribution in jet $E_T$ and from the invariant mass of the charged lepton and *b*-jet, where the error was 12% compared with the current one of 7%. Also, the older all-hadronic result improved from a 6.2% systematic error compared with the new value of 3.0%. The improvement in the all-hadronic channel was obtained from a better understanding of gluon radiation in the model of $t\bar{t}$ production, and from changes in mass fitting. The best values of top quark mass are from the lepton+jets channel in each of the experiments.

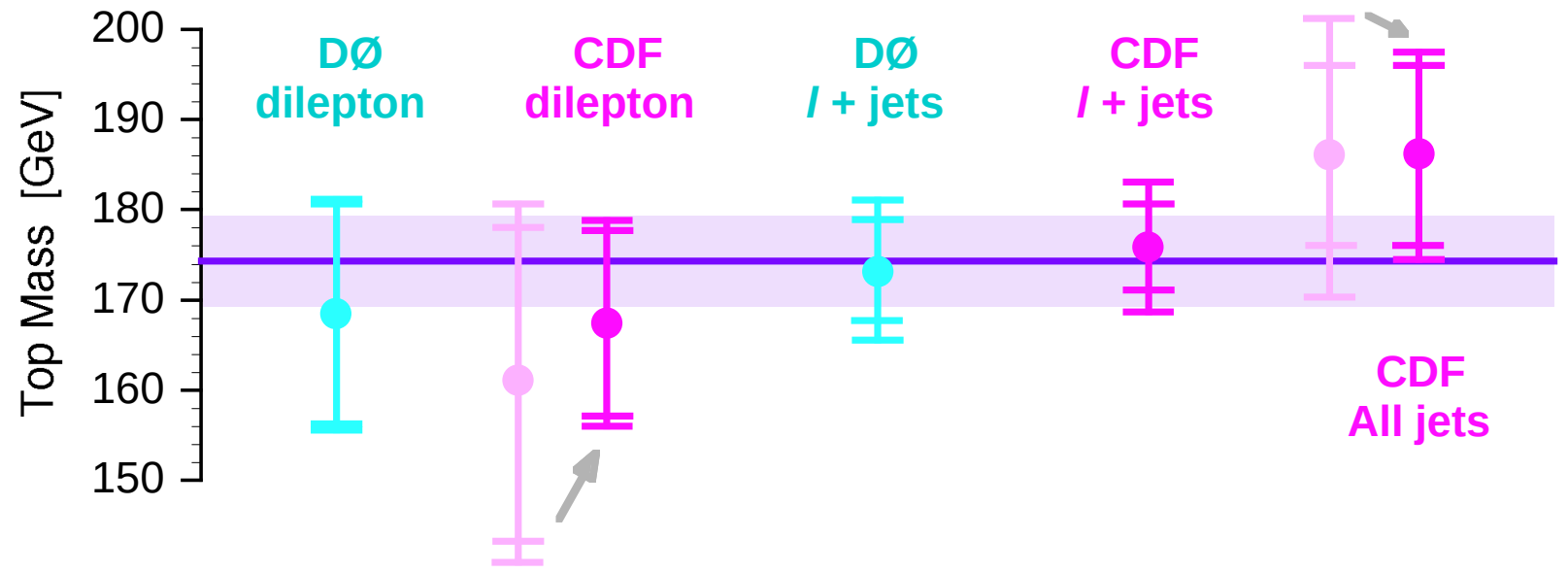

Figure 5 The five measurements (points and error bars) and the average result (line and error band) for the combined Tevatron top quark mass.

The systematic errors on the five mass measurements were each divided into six components with 22 subparts, and the correlations between the measurements were determined for each component separately. The resulting top quark mass is:

Tevatron $$m_t = 174.3 \pm 3.2 \text{ (stat)} \pm 4.0 \text{ (syst) GeV} = 174.3 \pm 5.1 \text{ GeV}$$

Figure 6 shows how the measurements contribute to the central value of the mass when the correlations between components are taken into account (first pie chart), when correlations between channels are ignored (second pie chart), and the 3-d bar chart shows the correlations between the channels. Most of the correlation is indirectly related to jet multiplicity via the uncertainty on the jet energy scale, since that forms the largest part of the systematic error on each of the measurements (48% for DØ dilepton to 88% for CDF all-hadronic).

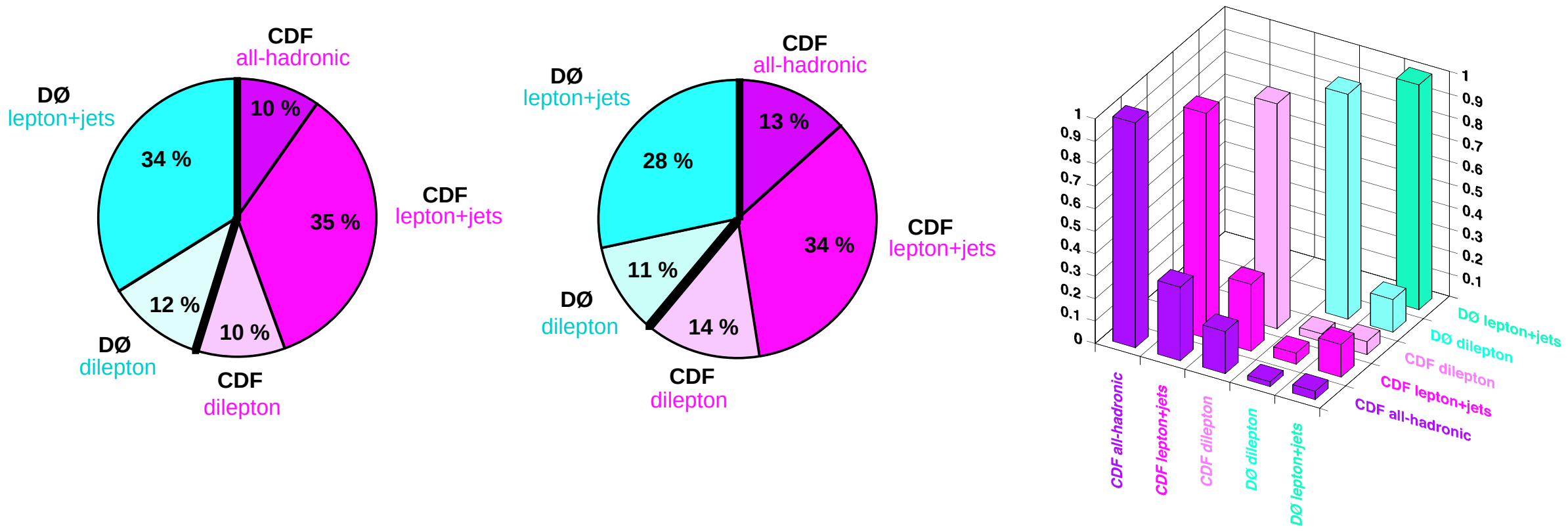

Figure 6 The first pie chart shows the relative contributions of each measured top quark mass to the combined central value, when correlations between the measurements are taken into account. The second pie shows the contributions when the correlations are ignored. The 3-d bar chart illustrates the correlations between the various measurements.

## 4. *W* Boson Helicity in Top Quark Decay

In the standard model, top decays only to a longitudinally polarized (helicity $h_W = 0$) or left-handed ($h_W = -1$) *W* boson, and not to a right-handed one. At leading order, the decay ratio is:

$$\frac{B(t \to bW_{\text{long}})}{B(t \to bW_{\text{left}})} = \frac{1}{2}\left(\frac{m_t}{m_W}\right)^2 = \frac{0.70}{0.30}$$

Information about the helicity of the *W* boson can be obtained by measuring the direction of the charged lepton from the *W* decay. However, angular distributions are affected by combinatorics and by the ambiguity inherent in neutrino reconstruction, and another way of determining the helicity is to measure the transverse momentum of the charged lepton. CDF[8] has fit the lepton $p_T$ distribution in lepton+jets events and in dilepton events to the hypothesis that there are no right-handed decays, to determine the fraction of longitudinal *W* decays, and they have then allowed the fraction of right-handed decays to float in the fit while assuming the standard model fraction of longitudinal decays. Figure 7 shows the result of the fit when no right-handed component is allowed. The following preliminary results are obtained:

CDF $\quad B(t \to bW_{\text{long}}) = 0.97 \pm 0.37 \pm 0.12 \quad [\text{for } B(t \to bW_{\text{right}}) = 0.0]$

$\quad B(t \to bW_{\text{right}}) = 0.11 \pm 0.15 \pm 0.06 \quad [\text{for } B(t \to bW_{\text{long}}) = 0.7]$

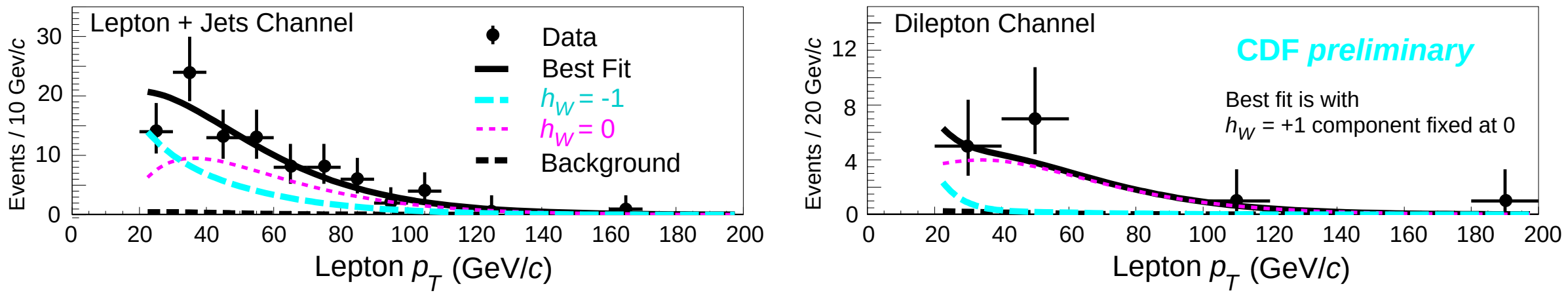

Figure 7 Fits of *W* boson helicity fraction contributions to the charged lepton $p_T$ distribution in top quark decays.

## 5. Top-Antitop Spin Correlations

The spins of the top quark and its antiquark are highly correlated in $t\bar{t}$ production in the standard model. One can maximize the expected correlation through a judicious choice of basis in which to analyze the events. For $p\bar{p} \to t\bar{t}$ at 1.8 TeV, about 90% of the events come from $q\bar{q}$ annihilation, with only 10% originating from gluon fusion, and the best basis to work in is the "off-diagonal" basis[9] optimized for $q\bar{q} \to t\bar{t}$. This is illustrated in Fig. 8, together with two other common bases.

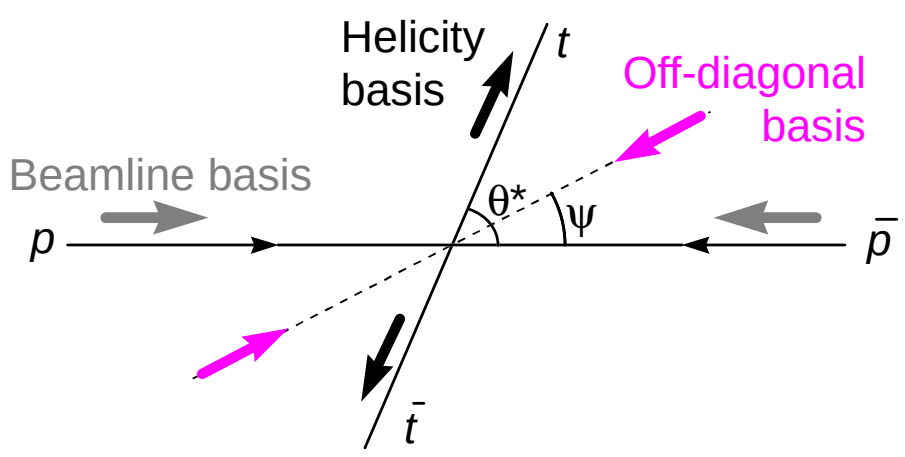

Figure 8 Illustration of the off-diagonal basis, defined by angle $\psi$ relative to the beamline, the helicity basis (defined by angle $\theta^*$) and the beamline basis. In the relativistic limit of the top quark's velocity, the off-diagonal basis tends to the helicity basis, and when the top is at rest, it tends to the beamline basis.

The angles $\psi$ and $\theta^*$ are related as follows: $\quad \tan\psi = \dfrac{\beta^{*2}\sin\theta^*\cos\theta^*}{1-\beta^{*2}\sin^2\theta^*}, \quad$ where $\beta^*$ is the velocity of the top quark in the constituent quark rest frame.

The differential decay rate of the top quark is related to the angle $\theta_\pm$ of the positive/negative lepton in the off-diagonal axis of the top quark rest frame as follows:

$$\frac{4}{\Gamma}\frac{d^2\Gamma}{d\cos\theta_+ d\cos\theta_-} = 1+\kappa\cos\theta_+\cos\theta_-$$

In the standard model, $\kappa$~0.9. The DØ collaboration[8] analyze their six dilepton events in the off-diagonal basis, and perform a binned maximum likelihood fit in the $(\cos\theta_+,\cos\theta_-)$ plane to obtain a lower limit on $\kappa$. (In Ref. 8, an alternative fitting method is illustrated. It yields the same result.) The preliminary results are shown in Fig. 9. The measured limit is:

DØ $\qquad \kappa > -0.2$ at the 68% confidence level.

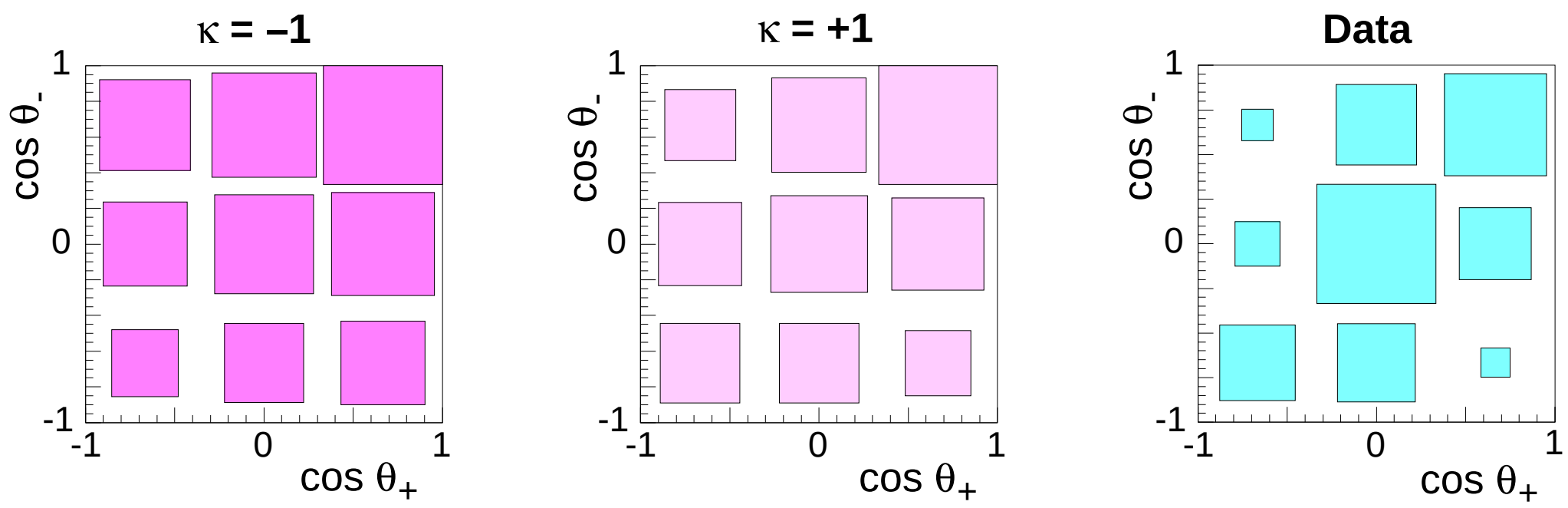

Figure 9 Distributions when the spins of the top and antitop are 100% anticorrelated, 100% correlated, and for data.

## 6. Invariant Mass of the $t\bar{t}$ System

It is interesting to examine the invariant mass spectrum of the $t\bar{t}$ system to search for evidence of a high mass resonance. Figure 10 shows example spectra from DØ[10] and CDF[11] for lepton+jets events. The DØ spectrum is from events used in the mass fitting, and shows a 2C fit without the top quark masses constrained to a specific value. If such a constraint is added, then the distribution of DØ events looks much like the one shown for CDF events, which had the constraint applied for $m_t = 175$ GeV. CDF has also cut on the fit $\chi^2$, and they demand 150 < $m_{3body}$ < 200 GeV, thereby reducing an excess at low $m_{t\bar{t}}$ caused by reconstructing top with a wrong jet combination. The cuts have been optimized to maximize the sensitivity to possible new physics. CDF is still working on including the systematic errors into the analysis, and they expect to set a limit on the mass of a top quark condensate $m_{Z'} > \sim 650$ GeV.

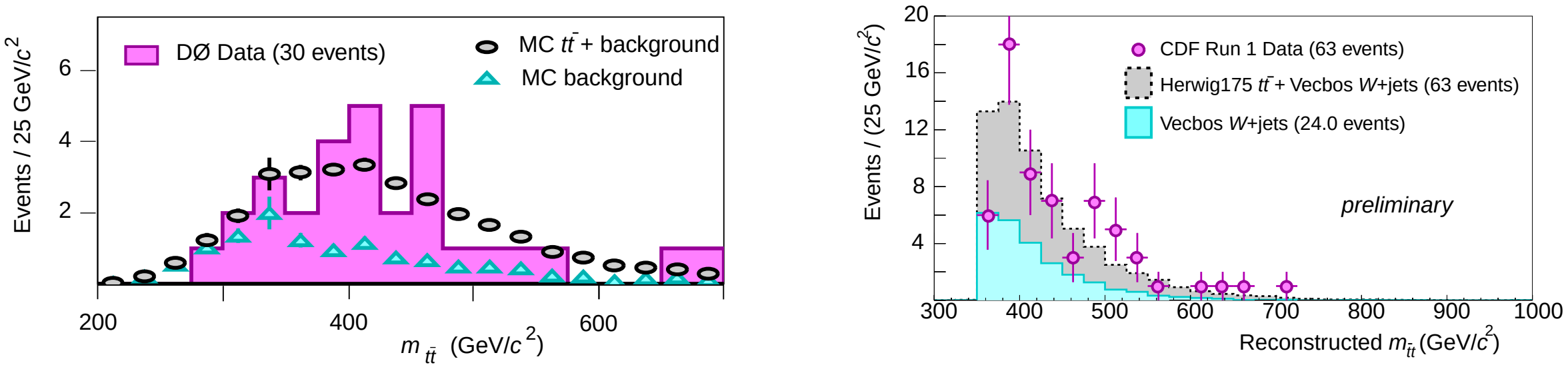

Figure 10 Invariant mass distributions of the $t\bar{t}$ system, for DØ on the left, and CDF on the right.

## 7. Top Quark Decay to a Charged Higgs Boson

In supersymmetric extensions of the standard model, the top quark can decay to a charged Higgs boson and a $b$ quark. If $\tan\beta$ is low, the Higgs will decay to $\bar{s}c$, and if it is high, the Higgs decays into $\tau\nu$. CDF[8] and DØ[12] have performed "disappearance" experiments in both regions, by seeing how much phase space is left for such decays after all reconstructed $Wb$ decays are taken into account. CDF has also performed a direct search for $\tau\nu$ production in top decays, which sets less severe constraints. The results are shown in Fig. 11.

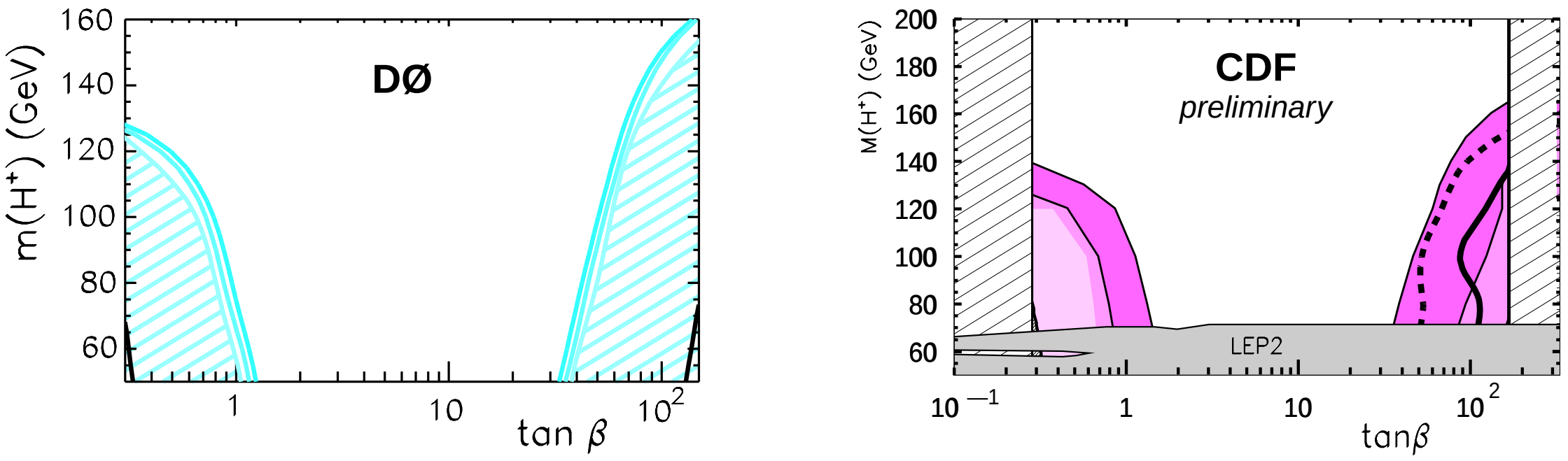

Figure 11 Exclusion regions in the $\left(\tan\beta, m_{H^+}\right)$ plane.

The hatched regions in the DØ plot, and solid shaded regions in the CDF plot are excluded at the 95% confidence level. The three lines at the limiting boundary in the DØ plot are for different values of the $t\bar{t}$ cross section, 5.5, 5.0, and 4.5 pb. The most conservative limit is from the measured value of 5.5 pb (the others are from resummed next-to-leading order calculations). On the CDF plot, the tightest limit is with $\sigma(t\bar{t})$ = 5.0 pb, the next tightest for 7.5 pb (close to the CDF measured value), and the least severe limit in the low $\tan\beta$ region is independent of the choice of $t\bar{t}$ cross section. In the high $\tan\beta$ region on the CDF plot, the solid and broken lines are results from the direct search with $\sigma(t\bar{t})$ = 5.0 and 7.5 pb, respectively.

The region $m_{H^+} <\sim 60$ GeV is excluded for all $\tan\beta$ by direct searches at the LEP experiments.[13] It is not possible to set limits in the regions off the DØ plot at very low and very high $\tan\beta$ (and also in the two small triangular regions in the lower corners of the plot), and for the lightly hatched regions on the CDF plot, because perturbative calculations are rendered invalid by the increasing strength of the Yukawa coupling of $H^+$ to $t$ and $b$ in these regions. In the (low $\tan\beta$, high $m_{H^+}$) region, there is no sensitivity for disappearance searches, because of the possibility of an alternate Higgs decay mode[14] of $H^+ \to t^* b \to Wbb$.

## 8. Electroweak Production of Single Top Quarks

The preceding measurements have all been associated with $t\bar{t}$ pair production. At the Tevatron collider, top quarks can also be produced singly, via the electroweak interaction. The CDF collaboration[11] has undertaken a search for single top events in the 109 pb$^{-1}$ of Run 1 data. Figure 12 shows the tree level Feynman diagrams for the two principal production channels.[15],[16]

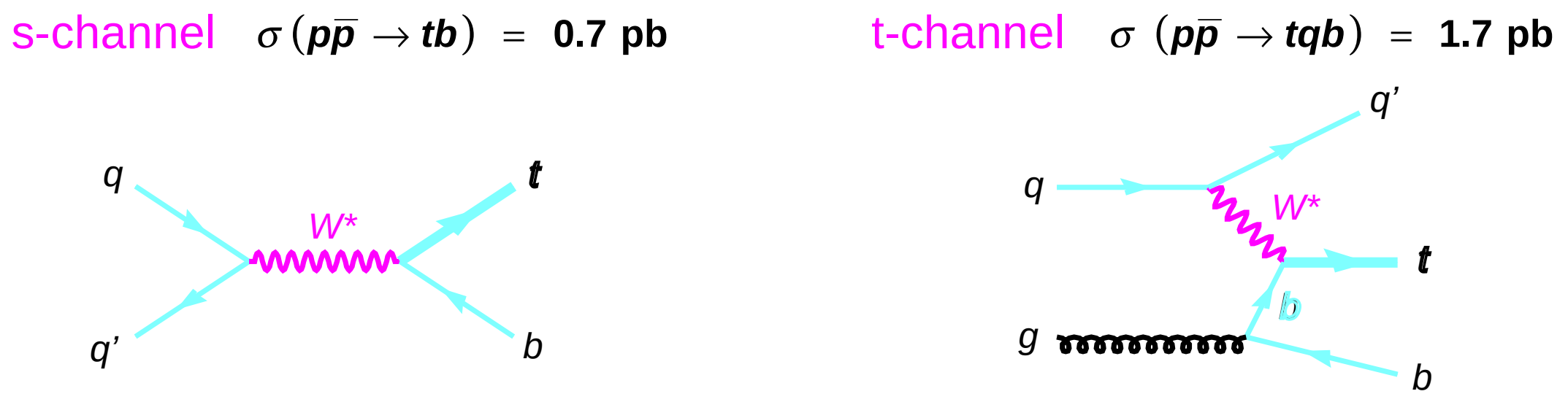

Figure 12 Leading order Feynman diagrams for $W^*$ single top production and $W$-gluon fusion at the Tevatron.

The search strategy is to identify events with one isolated electron or muon, significant missing transverse energy from the decay of the $W$ boson, exactly two jets, and one or more SVX $b$-tagged jets for the s-channel, or exactly one SVX $b$-tagged jet in the t-channel and an invariant mass window requirement of $145 \text{ GeV} < m_t < 205 \text{ GeV}$.

Figure 13 shows the distributions of background, sum of signal and background, and data, for each of the production modes. In the t-channel, the quantity "lepton charge × untagged jet pseudorapidity" takes advantage of the forward (backward) direction of the light quark produced with the top (antitop).

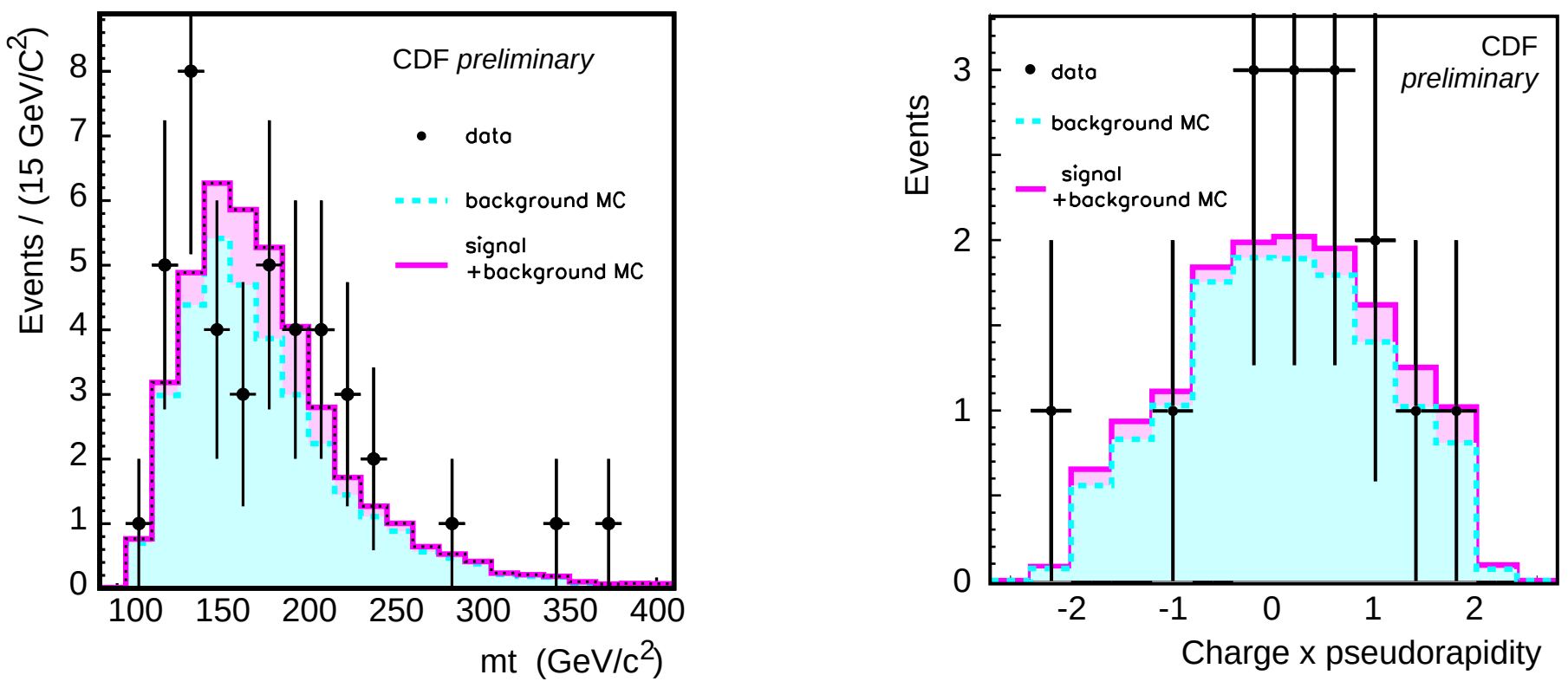

Figure 13 The left plot shows the reconstructed top mass distribution for s-channel single top production, and the right plot shows lepton charge times untagged jet pseudorapidity for t-channel single top production.

From likelihood fits to these distributions, the following upper limits on the cross sections are obtained:

CDF

$$\sigma\left(p\bar{p} \to t\bar{b},\ \bar{t}b\right) < 15.8 \text{ pb} \quad \text{at the 95\% confidence level}$$

$$\sigma\left(p\bar{p} \to tq\bar{b},\ \bar{t}\bar{q}b\right) < 15.4 \text{ pb} \quad \text{at the 95\% confidence level}$$

The s-channel limit is $22\times$ higher than the predicted NLO cross section, whereas the t-channel limit is only $9\times$ higher than that prediction. Such searches should be successful using the new data expected from Tevatron Run 2, scheduled to start next year.

## 9. Summary

Top quark physics at the Fermilab Tevatron is an active and flourishing field. A wide variety of measurements have now been made, and preparations are underway for improving them with significantly upgraded detectors and much-increased statistics in the next few years.

## Acknowledgments

I would like to thank the organizers of the XXXIVth Moriond Meeting on Electroweak Interactions and Unified Theories, especially J. Trân Thanh Vân, for hosting a most interesting and enjoyable meeting. I would also like to thank the members of the DØ and CDF collaborations who helped me prepare material for the talk and this paper.